# Similarity and delay between two non-narrow-band time signals

Zhen Sun[a], Guocheng Wang[a], Xiaoqing Su[b], Xinghui Liang[a] and Lintao Liu*[a]

[a]State Key Laboratory of Geodesy and Geodynamics, Academy of Precise Measurement Science and Technology, Chinese Academy of Sciences, Wuhan, 430077
[b]School of Civil and Architectural Engineering, Shandong University of Technology, Zibo, 255049

**Abstract**: Correlation coefficient is usually used to measure the correlation degree between two time signals. However, its performance will drop or even fail if the signals are noised. Based on the time-frequency phase spectrum (TFPS) provided by normal time-frequency transform (NTFT), similarity coefficient is proposed to measure the similarity between two non-narrow-band time signals, even if the signals are noised. The basic idea of the similarity coefficient is to translate the interest part of signal $f_1(t)$'s TFPS along the time axis to couple with signal $f_2(t)$'s TFPS. Such coupling would generate a maximum if $f_1(t)$ and $f_2(t)$ are really similar to each other in time-frequency structure. The maximum, if normalized, is called similarity coefficient. The location of the maximum indicates the time delay between $f_1(t)$ and $f_2(t)$. Numerical results show that the similarity coefficient is better than the correlation coefficient in measuring the correlation degree between two noised signals. Precision and accuracy of the time delay estimation (TDE) based on the similarity analysis are much better than those based on cross-correlation (CC) method and generalized CC (GCC) method under low SNR.

## 1. Introduction

Correlation coefficient, ranging within [-1,1], can measure the degree to which two signals are related to each other. It was first proposed by Carl Pearson[1]. It is an important index in the correlation analysis. However, the correlation coefficient is susceptible to the effects of noises on the signals. To reduce the effects of noises on correlation measurement forms the motive of this study.

By measuring the correlation between two signals, we can estimate the time delay between two similar signals received at two spatially separated sensors in the fields of sonar, radar, biomedicine, geophysics, etc. [2-5]. One of the applications has been found in a global positioning system where the location of a radiating object can be determined using differential satellite path delay measurements [6]. Currently, most of the time delay estimation (TDE) algorithms including cross-correlation (CC) method, generalized cross-correlation (GCC) method, and generalized phase spectrum (GPS) method, etc. are based on the correlation analysis that is weak in filtering ability [7-8]. For example, the CC method, being the simplest correlation analysis method for TDE, is easy to be affected by noises, which severely limits its applications [9].

The GCC method introduces pre-filter such as the Roth filter, the smoothed coherence transform, the phase transform, etc. to reduce the effects of white Gaussian noise on the TDE before taking cross-correlation [10]. It improves a lot in solving the noises problem of TDE. The TDE variance can reach Cramer-Rao Lower Bound by using maximum likelihood (ML) approach[11]. However, the GCC method requires estimation of both source and noise spectra, which often gives rise to a large delay variance, particularly for short data lengths [12-13]. Besides, the true location of the maximum correlation coefficient is not constrained to discrete increments, and it may fall between the discrete sampling points on account of the sampling in time that results in estimation inaccuracy for the CC method and GCC method. An interpolation technique usually is used to improve the TDE, hence the complexity will increase [14]. The delay time resolution of the GPS method can be smaller than the adoption interval [8], however, when the signal power spectrum curve has considerable fluctuation, the accuracy of TDE will decrease significantly.

A lot of computational results show that those methods have poor results and even fail to solve the TDE when processing non-stationary signals under low signal-to-noise ratios (SNR). Wavelet transform and Hilbert-Huang transform have recently been shown to be useful mathematical tools to improve the accuracy of TDE in applications such as multi-path delay and variable delay of non-stationary signals[15-19]. However, the wavelet base is not unique, and choosing a different wavelet base will have a great effect on TDE. Additionally, the Hilbert-Huang transform has terminal flying wings in the EMD decomposition process, which causes false model function components and affect the accuracy of TDE.

Considering that the correlation coefficient is susceptible to the effects of noises, Lintao Liu proposed the concept of similarity coefficient between two time signals at a summer seminar of Institute of Geodesy and Geophysics, Chinese Academy of Sciences in 2019. Such concept can measure the similarity degree and time delay between two time signals, even if the signals are noised. The similarity coefficient is based on the time-frequency phase spectrum (TFPS) provided by normal time-frequency transform (NTFT). It directly utilize the instantaneous phases and instantaneous amplitudes of the NTFT, not using the inverse transform. Because the NTFT provides a platform for filtering, the similarity coefficient can reduce the effects of noises on the signals to be measured. This present paper will evaluate the performance of the similarity analysis under different signal-to-noise ratios (SNR).

The structure of the paper is as follows: Section two introduces the NTFT and defines the similarity coefficient and time delay between two time signals. Section three gives some simulation experiments, showing that the similarity coefficient works better than the correlation coefficient when the signals are noised and that the similarity analysis performs better than the CC method and GCC method in the TDE. Some conclusions are given in Section four.

## 2. Definition of Similarity coefficient

### 2.1 NTFT definition

NTFT[20-23] is the premise for the definition of similarity coefficient. Here we briefly introduce the NTFT.

*Definition 1*: For a time function $f(t) \in C$, its linear time-frequency transform can be defined as

$$\Psi f(\tau,\varpi) = \int_{-\infty}^{+\infty} f(t)\bar{\psi}(t-\tau,\varpi)dt \qquad \varpi,\tau \in \boldsymbol{R} \tag{1}$$

where, $\tau$ is the time index, $\varpi$ is the frequency index, $\psi$ represents the transform kernel, the overline "—" denotes the conjugate operator, $\boldsymbol{C}$ is the complex field and $\boldsymbol{R}$ the real filed. The equation (1) is called a *NTFT* if kernel function's Fourier transform

$$\hat{\psi}(\omega,\varpi) = \int_{-\infty}^{+\infty} \psi(t,\varpi)\exp(-i\omega t)dt \tag{2}$$

satisfies:

$$\hat{\psi}(\omega,\varpi) = 1 \quad \text{when} \quad \omega = \varpi \tag{3}$$

$$|\hat{\psi}(\omega,\varpi)| < 1 \quad \text{when} \quad \omega \neq \varpi \tag{4}$$

where, the hat " ˆ " denotes the Fourier transform operator, $|\bullet|$ refers to the modulus operator.

A typical kernel function of NTFT can be designed as:

$$\psi(t,\varpi) = |\mu(\varpi)| w(u(\varpi)t) \exp(i\varpi t) \qquad w(t) \in \Omega(R) \tag{5}$$

where, $\mu(\varpi)$ is rescaling factor, determining the type of the NTFT. For example, letting $\mu(\varpi) = 1$ the NTFT becomes a normal Gabor transform and letting $\mu(\varpi) = \varpi$ the NTFT becomes a normal wavelet transform. $\Omega(R)$ denotes the set of normal windows. $w(t)$ is called a *normal window* if its Fourier transform satisfies

$$\hat{w}(\omega) = |\hat{w}(\omega)| = \text{Maximum} = 1 \Leftrightarrow \omega = 0 \tag{6}$$

The normal Gaussian window is

$$w(t) = \frac{1}{\sqrt{2\pi}\sigma} \exp(-\frac{t^2}{2\sigma^2}) \tag{7}$$

Where, $\sigma$ is a positive constant. In this study, we adopt the normal Morlet wavelet transform (NMWT) in measuring the similarity between two time signals. The NMWT is a special NTFT that sets $\mu(\varpi) = \varpi$ and adopts the normal Gaussian window (as Eq.7).

For a harmonic $h(t) = A\exp[i(\omega t + \phi)]$, its NTFT satisfies

$$|\Psi h(\tau,\varpi)| = \text{Maximum} \Leftrightarrow \varpi = \omega, \forall \tau \in R \tag{8}$$

$$\Psi h(\tau,\omega) = h(\tau), \forall \tau \in R \tag{9}$$

where $A$ is the amplitude, $\omega$ is the angle frequency, $\phi$ is the initial phase and $\forall$ means "for any". The properties as shown in equation (8) and equation (9) are called the *Inaction principle*. It guarantees that the NTFT can directly and unbiasedly reveal the immediate frequency, immediate phase and immediate amplitude of a harmonic [24-25].

**2.2 similarity coefficient**

*Definition 2:* For two time functions $f_1(t) \in R$ and $f_2(t) \in R$, their similarity function is defined as

$$\rho(s) = \frac{\iint_{S+s} \text{Re}\,\Psi f_1(\tau-s,\varpi) \text{Re}\,\Psi f_2(\tau,\varpi) d\tau d\varpi}{\sqrt{\iint_S (\text{Re}\,\Psi f_1(\tau,\varpi))^2 d\tau d\varpi \iint_{S+s} (\text{Re}\,\Psi f_2(\tau,\varpi))^2 d\tau d\varpi}} \tag{10}$$

where, $s \in R$ is translating length, $S$ is the interest time-frequency area of $f_1(t)$; $S+s$ is the $S$ area translated along the time axis by $s$, Re denotes the real part. The maximum of the similarity function is called as *similarity coefficient*. The location of the maximum indicates the *time delay* between $f_1(t)$ and $f_2(t)$.

An equivalent version of above similarity function is

$$\rho(s) = \frac{\iint\limits_S \operatorname{Re}\Psi f_1(\tau,\varpi)\operatorname{Re}\Psi f_2(\tau+s,\varpi)d\tau d\varpi}{\sqrt{\iint\limits_S (\operatorname{Re}\Psi f_1(\tau,\varpi))^2 d\tau d\varpi \iint\limits_S (\operatorname{Re}\Psi f_2(\tau+s,\varpi))^2 d\tau d\varpi}} \quad (11)$$

The version of similarity function is easy to understand and code.

There are several points to be made about the similarity coefficient. Firstly, we utilize the time-frequency phase spectrum (TFPS) of NTFT. For a real time signal, the real part of its NTFT is its TFPS. One can translate the interest part of $f_1(t)$'s TFPS along the time axis to couple with the $f_2(t)$'s TFPS. Such coupling should generate a maximum if $f_1(t)$ and $f_2(t)$ are really similar to each other in time-frequency structure. The location of the the maximum should indicate the time delay between two similar signals.

Secondly, $f_1(t)$ and $f_2(t)$ should have certain frequency band width. So the similarity coefficient defined here only applies to two non-narrow-band time signals. We will define the similarity and delay between two narrow-band time signals in another paper.

Thirdly, the similarity coefficient definition refers to the correlation coefficient definition. Both of coefficients are in nature an outcome of coupling. The difference is that the coupling in similarity coefficient is two-dimensional while the coupling in correlation coefficient is one-dimensional. Like the correlation coefficient, the similarity coefficient ranges within [-1,1]. However, the similarity coefficient disuses the concept of expectation, because a real signal $f(t)$, if being oscillating, should satisfy

$$\iint\limits_S \operatorname{Re}\psi f(\tau,\varpi)d\tau d\varpi \approx 0 \quad (12)$$

In other words, the expectation of the TFPS of a real oscillating signal is nearly zero. To say more, the phase structure of the NTFT would be violated if introducing the expectation concept in similarity coefficient.

Lastly, the NTFT provides a platform for time-frequency filtering, that is why the similarity coefficient can resist the noises. Furthermore, the similarity coefficient disuses the inverse transform of NTFT, which means it has the advantage of directness.

**2.4 Origin of similarity coefficient**

The similarity coefficient originates in the study on the shock signal of the Tianjin Binhai New Area explosion accident on August 12, 2015. The accident includes two consecutive explosions. The total energy of the explosion was approximately 450 tons of

TNT equivalent. NTFT spectrum of the explosion shock signal shows that the two explosion shocks, though being very different in significance, are very similar to each other in time-frequency structure (particularly in TFPS) (Fig.1).

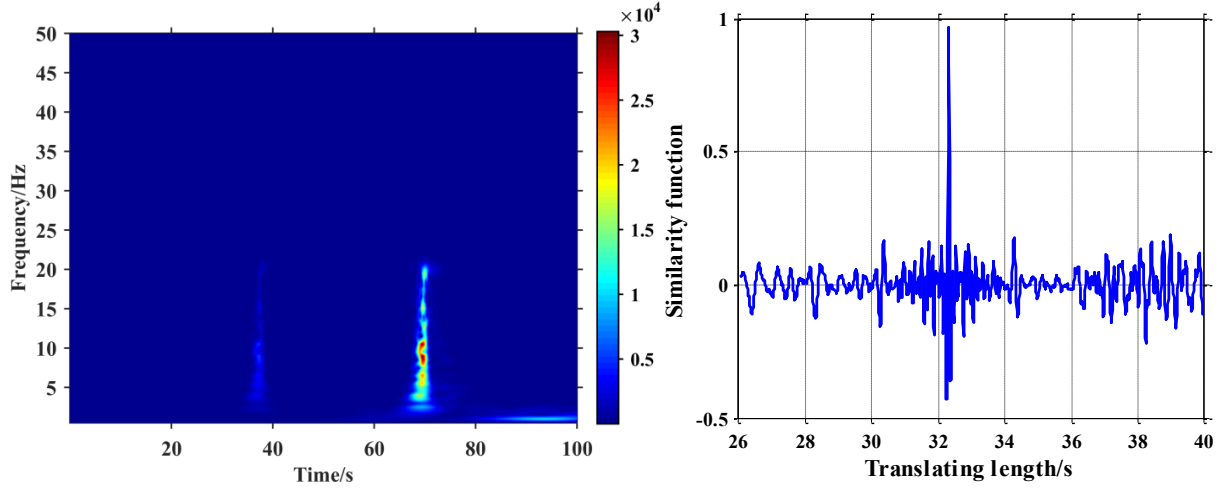

Fig.1 NTFT spectrum of 2015 Tianjin explosion shock signal (left) and similarity function of the two shocks.

The similarity coefficient reaches up to 96%, strongly suggesting that the two explosions take place at the same place with the same mode. However, the correlation coefficient of the two explosion shocks is only 78%, if using the correlation analysis. The obvious difference between similarity analysis and correlation analysis, resulting from some noises, reveals that the similarity analysis is more accurate and more sensitive in judging the correlation degree between two time signals. Furthermore, the similarity analysis can easily reveal that the time interval (i.e. delay) between two explosions is 32.31 seconds.

## 3. Simulated tests

In this section, we will simulate two non-narrow band pulse signals under different SNR to evaluate the performance of the similarity analysis. The two non-narrowband pulse signals have a length of 5.4s, and their sampling rate is 1000Hz:

$$\begin{cases} f_1(n) = \sin(2\pi\omega(n/f_s)^2) + \varepsilon_1 \\ f_2(n) = \sin(2\pi\omega((n-d)/f_s)^2) + \varepsilon_2 \end{cases} \quad (13)$$

where, $f_s$ is the sampling frequency; $n$ is the length of signal; $\omega = 5Hz$ is the frequency of two non-narrow-band pulse signals; $d$=150 is delay point, namely, the time delay between two signals is 0.15 seconds; $\varepsilon_1$ and $\varepsilon_2$ represent the stationary white noise, uncorrelated with the signals or each other.

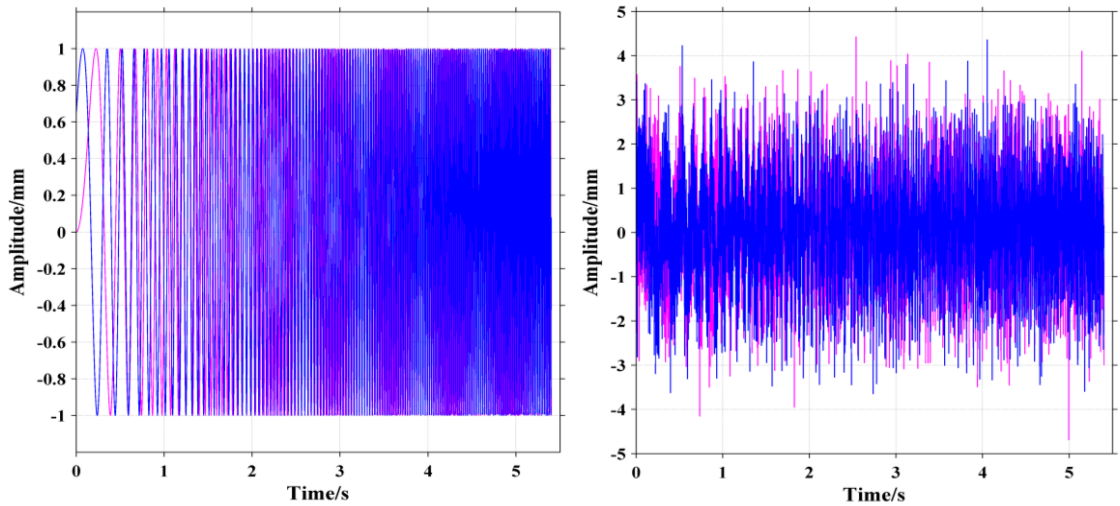

Fig.2 Two non-narrowband pulse signals without noise (left) and with noises(right).

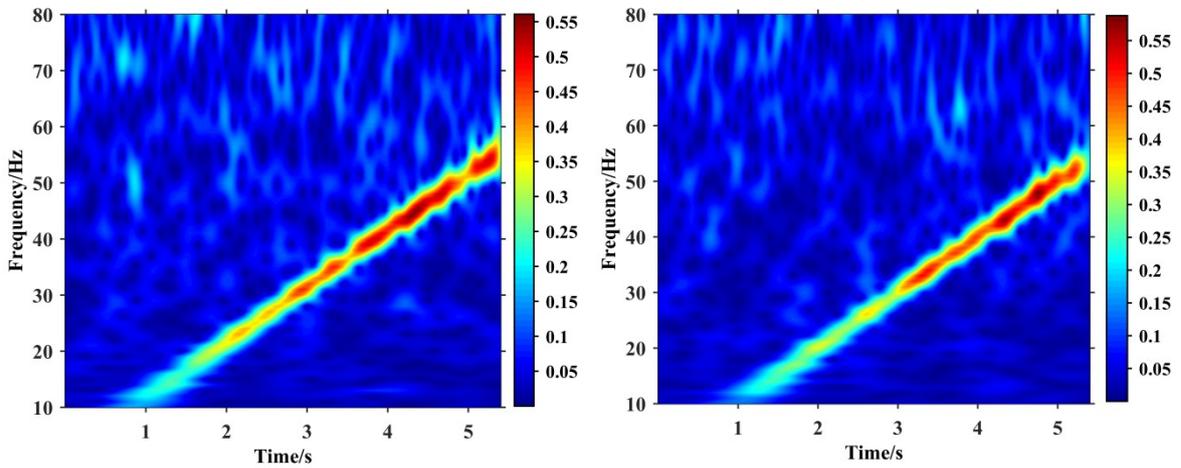

Fig.3 NTFT spectra of the two non-narrow-band pulse signals with noises.

In Fig.2, the left is the original time series of two non-narrow band pulse signals; the right represents the two time series disturbed by noises when the SNR is zero. The pink represents the first signal, and blue represents the second signal. One can see from Fig.2 (right) that the signals are heavily disturbed. Fig.3 shows the time-frequency spectra of the two noised signals in Fig.2(right). One can find that the NTFT has the capacity of filtering, and can reduce the impact of background noises.

We here design two simulated tests. The first aims to compare similarity coefficient with correlation coefficient in measuring the correlation degree of the two above signals disturbed by different noises. The second aims to show that the similarity analysis works better than the CC method and GCC method in the TDE.

**3.1 Measuring the similarity between two time signals**

We compute the similarity/correlation functions in cases of SNR = 0 dB and SNR =-10 dB. The results are shown in Fig.4 and Fig.5.

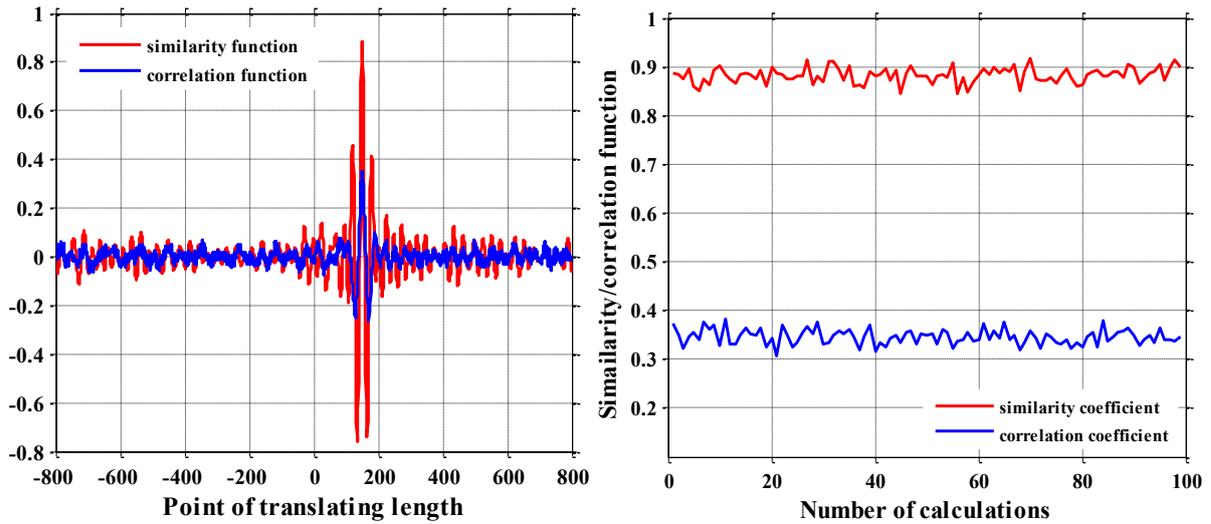

Fig.4 Comparison between similarity analysis and correlation analysis in case of SNR = 0 dB

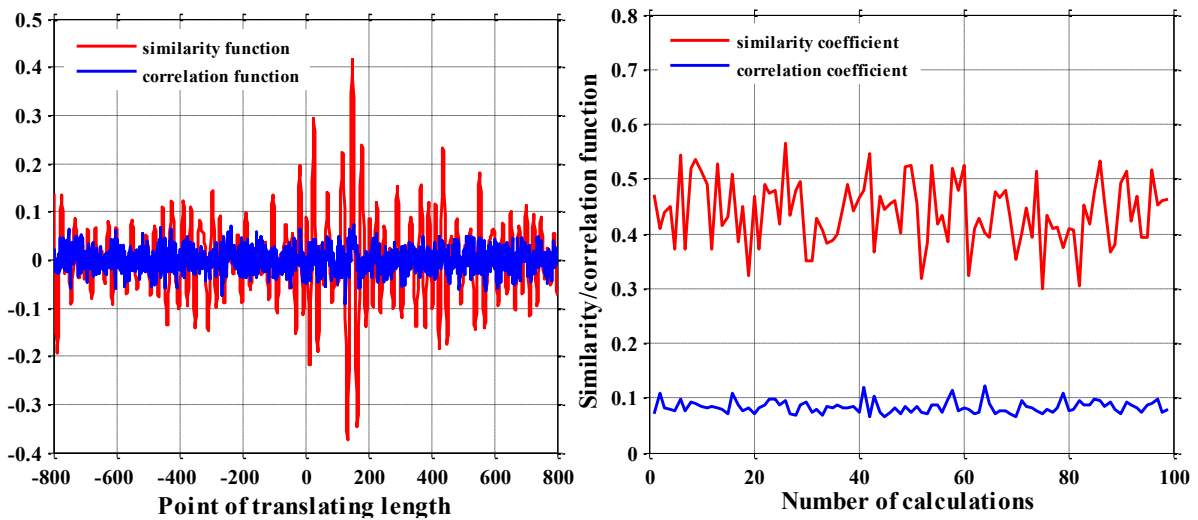

Fig.5 Comparison between similarity analysis and correlation analysis in case of SNR =-10 dB

In case of SNR=0dB, the correlation coefficient is only 0.3491, seeming to indicate that there is a week positive relationship between two non-narrow-band pulse signals. However, that is an inaccurate judgment. This means correlation coefficient is somewhat impractical. The similarity coefficient is 0.8850, manifesting that there is a very strong positive relationship between the two time signals though they are noised. Here, the filtering ability of the similarity coefficient is obvious.

We randomly calculate for 100 times to avoid the computational contingency under the same SNR. Each time can generate different Gaussian noises. As can be seen from fig. 4 (right), the average of the similarity coefficients is 0.8833. However, the average of the correlation coefficients is only 0.3446. These results show that the similarity coefficient is more effective than the correlation coefficient in measuring the correlation degree between the two noised signals.

As shown in Fig.5, the correlation coefficient cannot measure the correlation between the two time signals when the SNR is -10 dB. Moreover, the peak value of the traditional method has been seriously contaminated by the side peaks, which directly reduce the precision of TDE. The similarity coefficient can improve the accuracy of the correlation coefficient. Subsequently, the correlation/similarity coefficients are calculated 100 times. In each calculation, the Gaussian noises are different but the SNR remains the same. The results show that two noised signals have a moderate positive linear relationship by the similarity coefficient. However, the average of the correlation coefficient is below 0.1 due to the serious effect of background noises.

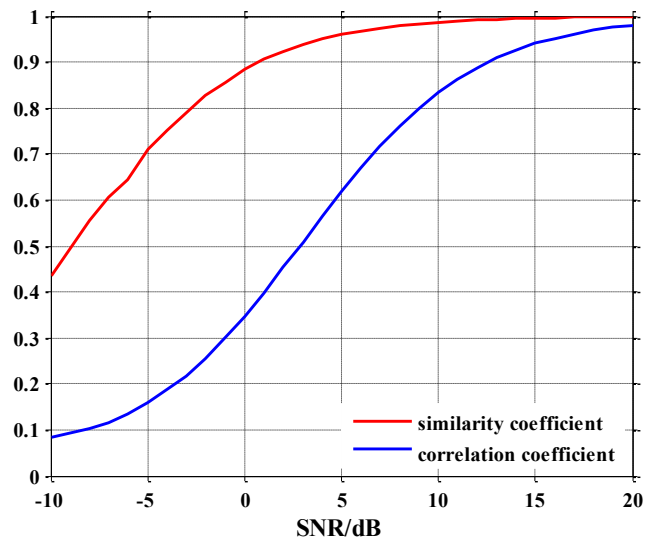

Fig.6 Similarity coefficient and correlation coefficient with respect to SNR.

Now, we measure the correlation degree of the two time signals under different SNR by using the correlation coefficient and similarity coefficient (Fig.6). Both methods can accurately determine the correlation degree when the two signals are slightly polluted or not affected by noise. However, the correlation coefficient can not measure the correlation degree between the two signals under low SNR. Meantime, its declining speed is faster than the similarity coefficient with the decrease of SNR. On the contrary, the similarity coefficient can more accurately measure the correlation degree between the two time signals under low SNR.

In summary, the similarity coefficient is more sensitive and accurate than the correlation coefficient when measuring the correlation degree between two noised time signals.

## 3.2 TDE

The location of the maximum of the similarity coefficient indicates the *time delay* between $f_1(t)$ and $f_2(t)$. Thus the similarity analysis can be used for the TDE. We mainly analyze the stability and accuracy of the TDE using the similarity analysis, comparing with the CC method and GCC method. The GCC method can mitigate the impact of measurement noise by selecting maximum likelihood pre-filters, which has been proven to be superior to the CC method at low SNR. This study adopts two indexes including success rate (SR) and the mean square error (MSE) to measure the TDE precision and accuracy. The calculation formula for success rate is

$$SR = \frac{\gamma}{n} \times 100\% \qquad (14)$$

where, $\gamma$ is the number of statistics when estimated solution $\tilde{c}$ from different TDE methods equals the true delay value $d$. $n$ is the number of calculations for the same SNR, i.e $n=100$. MSE is a measure of how close the solution result is to the true result. The calculation formula for MSE is

$$MSE = \sqrt{\sum_{i=1}^{n} \Delta_i^2 \Big/ n-1} \qquad (15)$$

where, $\Delta = d - \tilde{c}$ is the deviation of delay point.

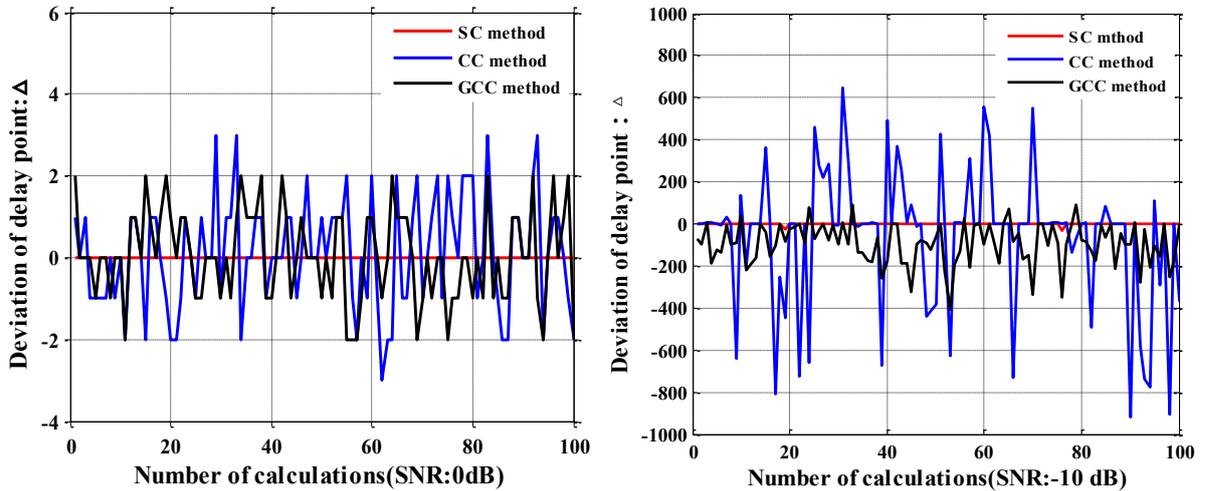

Fig.7 Deviation of the TDE by three methods

As shown been in Fig.7, TDE results of the CC method are extremely unstable because of the effect of observation noise when SNR is zero. The GCC method introduces pre-filter to reduce the influence of white Gaussian noise on the TDE before taking the CC method. It can improve the stability of the CC method algorithm's delay estimation results but still shows strong unstable characteristics. However, the method usually is introduced to deal with the time delay between two stationary signals, and its estimation performance falls in cases of solving non-stationary signals. The similarity coefficient (SC) method can stably and accurately determine the time delay between two non-stationary signals. The stability of its solution is better than the GCC method and CC method. The SC method improves the precision of the CC method and GCC method by one order of magnitude, and its success rate is as high as 100%, which is much higher than the CC method and GCC method. The performance of three methods falls in precision and accuracy when SNR drops to -10dB. The GCC method and CC method cannot give reliable TDE and their error rate is as high as 85% and 95%, respectively. The SC method can still estimate the time delay of the two time signals stably and its success rate is much higher than GCC method and CC method.

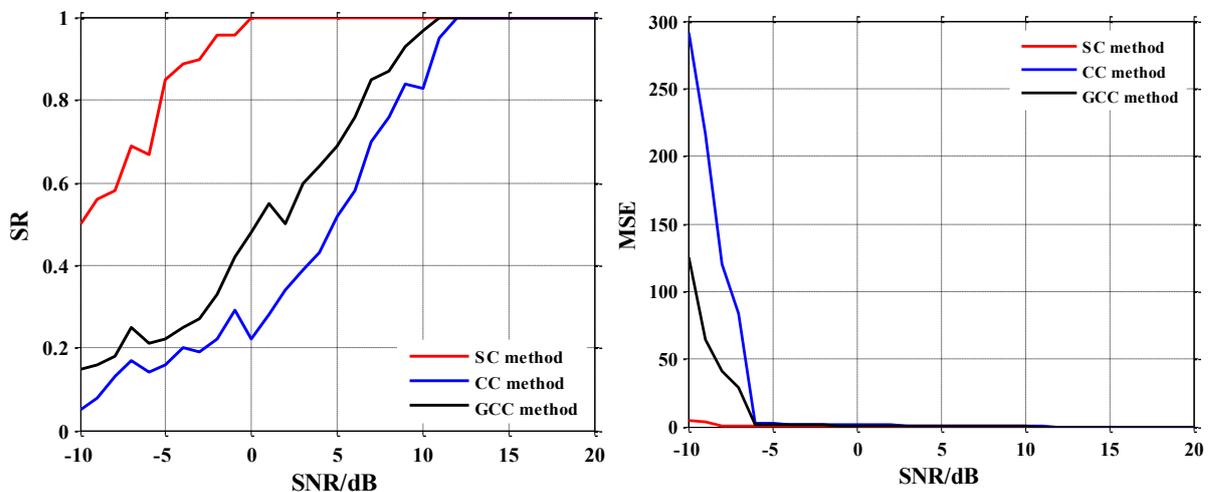

Fig.8 TDE results by three methods with respect to SNR.

Now, we estimate the time delay between the two signals with different SNR using the three methods. It can be seen from Fig.8 that the success rate of CC method and GCC method approaches to zero when SNR is -10dB, because the noises seriously disturb most of the useful components of the two signals. The GCC method can usually improve solution precision and calculation accuracy of the CC method through pre-filters. However, we will get worse results if we choose unsuitable pre-filters. It is hard to elect a suitable filter for different time series. Furthermore, the GCC method usually assumes the signals are stationary. That means it is not applicable to the TDE for non-stationary signals. The SC method that converts the time domain information of the two signals to the time-frequency domain can stably handle non-stationary signals, and can reduce the impact of noises through its filtering to enhance the precision and accuracy rate of TDE. As can be seen from Fig.8, the precision and success rate of SC method are always better than those of two other methods, particularly in low SNR cases. This test demonstrates the powerful performance of the similarity analysis in solving the TDE problem.

## 4. Summary

The correlation coefficient is difficult to measure correctly the correlation degree between two noised time signals. Similarity coefficient, having filtering ability, is proposed to accurately measure the correlation degree between two non-narrow-band time signals. The similarity coefficient is achieved by translating certain part of the TFPS of one signal along the time axis to couple with the TFPS of another signal. The location of the coupling maximum indicates the time delay between the two time signals. One simulated test shows that the similarity coefficient is more sensitive and accurate than the correlation coefficient in measuring the correlation degree between two noised signals. Another simulated test proves that similarity coefficient method performs much better (in both precision and accuracy) than the GCC method and CC method in estimating the time delay between two time signals with low SNR.

**Acknowledgment**

This study was supported by the National Natural Science Foundation of China (grantNos.41074050, 41704003), Ministry of Science and Technology of the People's Republic of China (grantNos. 2011YQ120045) and the Natural Science Foundation of Hubei Province, China (grant No.2019CFB795).